\documentclass[12pt,preprint]{aastex}
\slugcomment{accepted for publication in Astrophysical Journal Letters}
\shorttitle{Galactic \ion{O}{6} Emission in the Halo}
\shortauthors{B. Otte et al.}

\begin{document}
\title{{\em FUSE} Detection of Galactic \ion{O}{6} Emission in the Halo above
the Perseus Arm}
\author{Birgit Otte, W. Van Dyke Dixon, Ravi Sankrit}
\affil{Department of Physics and Astronomy, The Johns Hopkins University, 3400
North Charles Street, Baltimore, MD 21218}
\email{otte@pha.jhu.edu, wvd@pha.jhu.edu, ravi@pha.jhu.edu}

\clearpage

\begin{abstract}
Background observations obtained with the {\em Far Ultraviolet Spectroscopic
Explorer (FUSE)} toward $l=95\fdg4$, $b=36\fdg1$ show
\ion{O}{6}\,$\lambda\lambda$1032,1038 in emission. This sight line probes a
region of stronger-than-average soft X-ray emission in the direction of
high-velocity cloud Complex C above a part of the disk where H$\alpha$ filaments
rise into the halo. The \ion{O}{6} intensities,
$1600\pm300$\,photons\,s$^{-1}$\,cm$^{-2}$\,sr$^{-1}$ (1032\,\AA) and
$800\pm300$\,photons\,s$^{-1}$\,cm$^{-2}$\,sr$^{-1}$ (1038\,\AA), are the lowest
detected in emission in the Milky Way to date. A second sight line nearby
($l=99\fdg3$, $b=43\fdg3$) also shows \ion{O}{6}\,$\lambda$1032 emission, but
with too low a signal-to-noise ratio to obtain reliable measurements. The
measured intensities, velocities, and FWHMs of the \ion{O}{6} doublet and the
\ion{C}{2}$^\ast$ line at 1037\,\AA\ are consistent with a model in which the
observed emission is produced in the Galactic halo by hot gas ejected by
supernovae in the Perseus arm. An association of the observed gas with Complex C
appears unlikely.
\end{abstract}



\section{INTRODUCTION}

The interstellar medium (ISM) consists of several phases. The hottest phase
reaches temperatures of a few million degrees and is observed in X-ray emission.
Gas at slightly cooler temperatures ($\sim3\times 10^5$\,K) is best traced by
\ion{O}{6}, which has a strong resonance doublet at 1032/1038\,\AA. The
\ion{O}{6} ion is generally assumed to trace collisionally-ionized gas, because
photoionization by hot stars cannot easily create such high ionization states in
the diffuse ISM (e.g., Martin \& Bowyer 1990). Possible scenarios for the
production of \ion{O}{6} include the cooling of gas originally heated by
supernova shocks and turbulent mixing of hot and cold gas at the surfaces of
infalling high-velocity clouds (HVCs) as they pass through the Galactic corona.

To date, Galactic \ion{O}{6}\,$\lambda$1032 has been observed in emission along
six sight lines with the {\em Far Ultraviolet Spectroscopic Explorer (FUSE)}
(Dixon et al. 2001; Shelton et al. 2001; Shelton 2002a; Welsh et al. 2002). The
location of the \ion{O}{6} emitting gas along these sight lines is generally
unknown. \citet{wel} note the narrow range of the observed intensities and
suggest that the \ion{O}{6} emission originates in the boundary of the Local
Bubble. Absorption-line measurements along about 100 sight lines \citep{sav02}
reveal only a poor correlation of \ion{O}{6} with other tracers of the ISM, such
as neutral and ionized hydrogen or soft X-ray (SXR) emission. \citet{sav02}
conclude that the observed \ion{O}{6} absorption is best explained by a patchy,
absorbing disk with a scale height of about 2.3\,kpc. This thick disk corotates
with the Galactic plane and shows signs of outflow along about 20\,\% of the
observed sight lines.

We have observed \ion{O}{6} emission along two sight lines, only one of which
allows reliable measurements using line profile fitting. We compare the physical
parameters of the \ion{O}{6} emission lines along this sight line with gaseous
features observed at other wavelengths representing different phases of the ISM
to determine the location of the \ion{O}{6} emitting gas and to investigate
possible processes leading to the observed \ion{O}{6} emission.

\section{OBSERVATIONS AND DATA REDUCTION}

Our data sets (program IDs S4054801 and S4056101, hereafter SL5 and SL6,
respectively) were obtained with {\em FUSE} as background observations recorded
in time-tag mode on 2002 January 19--22. Our data are the first observations
taken after the recovery of the satellite from a reaction wheel failure in
December 2001 \citep{ake}. We used the data obtained through the low-resolution
(LWRS) aperture ($30\arcsec\times30\arcsec$) on detecter segment LiF1A (see
\citet{moos} and \citet{sahn} for a complete description of {\em FUSE} and its
on-orbit performance). The total exposure time of the 55 exposures of SL5 was
217\,ksec, with 95\,ksec recorded during orbital night. The eight exposures of
SL6 yielded a total exposure time of 32\,ksec, with 13\,ksec during orbital
night. During these observations, the satellite was pointed near the orbital
pole (see Table \ref{obs} for observational data).

\begin{figure}
\epsscale{0.5}
\plotone{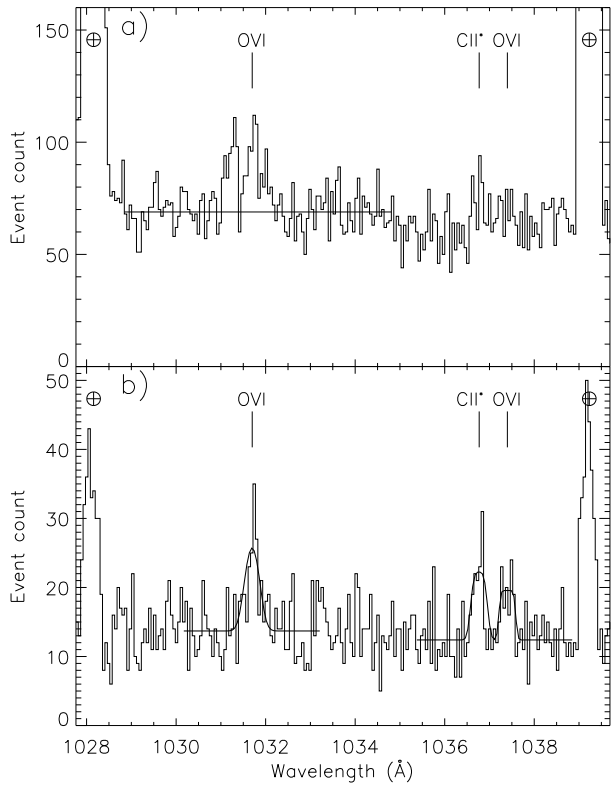}
\epsscale{1.0}
\caption{\label{spec} The SL5 spectra around the \ion{O}{6} doublet. Both
day-plus-night (a) and night only spectra (b) are shown. Both spectra were
binned by 8 pixels for display only. Panel a shows the continuum fit for
\ion{O}{6}\,$\lambda$1032; panel b shows the minimum $\chi^2$ fits for the
emission lines in SL5. The fitted positions of the \ion{O}{6} doublet and
the \ion{C}{2}$^\ast$ line are labeled. Airglow lines are marked with the Earth
symbol.}
\end{figure}

The data were processed using the CALFUSE pipeline version 2.0.5. Only the first
steps of the pipeline were executed, i.e. the data were screened for pulse
height (using limits 4--15) and corrected for Doppler effects of the satellite's
motion. Bursts were also removed, but no walk or drift correction was applied.
Then the data were combined using the program ttag\_combine. Nearby airglow
lines were used to determine the constant height of the extraction window. The
wavelength calibration was derived from the position of the airglow lines. The
resulting spectra for sight line SL5 are shown in Figure \ref{spec}. Lines of
\ion{O}{6}\,$\lambda\lambda$1032,1038 and \ion{C}{2}$^\ast\,\lambda$1037 are
clearly seen in the SL5 data. Due to the lower signal-to-noise ratio in the SL6
spectra, we will focus on the SL5 measurements in our analysis and refer only
briefly to the SL6 data in section \ref{sl6}.

We followed the arguments of \citet{sh01} to exclude terrestrial airglow, solar
\ion{O}{6} emission, and scattered light inside the satellite as other possible
explanations for the observed emission lines. Possible contamination (1$\sigma$
upper limit) of \ion{O}{6}\,$\lambda$1032 by fluorescing H$_2$ is less than
200\,LU (LU\,=\,photons\,s$^{-1}$\,cm$^{-2}$\,sr$^{-1}$) for SL5. An unknown
feature appears in the day-plus-night spectra on the blue side of
\ion{O}{6}\,$\lambda$1032 but is absent in the night-only data. This feature is
assumed to be airglow and has been seen in other data sets (e.g., Shelton et al.
2001). Due to its proximity to the \ion{O}{6}\,$\lambda$1032 line, reliable
measurements for \ion{O}{6} could be obtained only from orbital night-time data.

\section{\label{sl6}RESULTS}

We measured the line fluxes for SL5 in two ways. In the photon-counting method,
we summed the counts in the wavelength region occupied by the emission line and
subtracted a constant continuum derived from the surrounding spectrum. One-sigma
uncertainties were derived assuming Gaussian statistics. In the second method,
we used a line fitting algorithm, which determined the minimum $\chi^2$ for a
range of synthetic line profiles using three free parameters for each line:
central wavelength, line width (FWHM), and peak intensity. The background was
fitted by a straight line. We assumed that the emission line profiles were a
convolution of a Gaussian with the instrumental line spread function, which for
filled LWRS slit emission is a 106\,km\,s$^{-1}$ wide flat-top profile.
One-sigma uncertainties for each parameter were derived by varying the parameter
and optimizing $\chi^2$ with the other parameters until the newly calculated
$\chi^2$ had increased by 1 relative to the optimal solution \citep{avni}.

The results of both methods are given in Table \ref{meas} for the SL5 data.
$I_{\rm tot}$ is the integrated intensity of the emission line, FWHM is the
intrinsic line width, and $\lambda_{\rm cen}$ is the central wavelength of the
line. The derived velocity was converted to the local standard of rest ($v_{\rm
LSR}$). The large uncertainties of $I_{\rm tot}$ for \ion{O}{6}\,$\lambda$1038
and \ion{C}{2}$^\ast\lambda$1037 in the $\chi^2$ method reflect the close
proximity of these features: reducing the strength of one line was compensated
by increasing the strength of the other, allowing a larger parameter space with
only slow changes in $\chi^2$. The extinction toward SL5 is approximately
$E_{\rm B-V}=0.024$ (Schlegel, Finkbeiner, \& Davis 1998), corresponding to an
attenuation in the \ion{O}{6} intensity of $\sim30$\,\%, if all of the reddening
occurs in front of the \ion{O}{6} emitting gas and the extinction
parameterization of Cardelli, Clayton, \& Mathis (1989) is assumed. The line
ratio \ion{O}{6}\,$\lambda$1032/\ion{O}{6}\,$\lambda$1038 would approach 2 in
optically thin gas, whereas in an optically thick medium the line ratio would
approach 1 due to self-absorption. We derived a line ratio of $2.0\pm0.8$; the
large uncertainty makes it impossible to constrain the optical depth of the
\ion{O}{6} emitting gas.

Our second sight line, SL6, lies near SL5 and shows \ion{O}{6}\,$\lambda$1032
emission in the day-plus-night spectrum. However, the emission line is only
marginally detected in the night-only data and does not allow reliable
measurements. The day-plus-night spectrum suggests an \ion{O}{6}\,$\lambda$1032
intensity and velocity comparable to those measured in the SL5 data.

\section{DISCUSSION}

\begin{figure}
\plotone{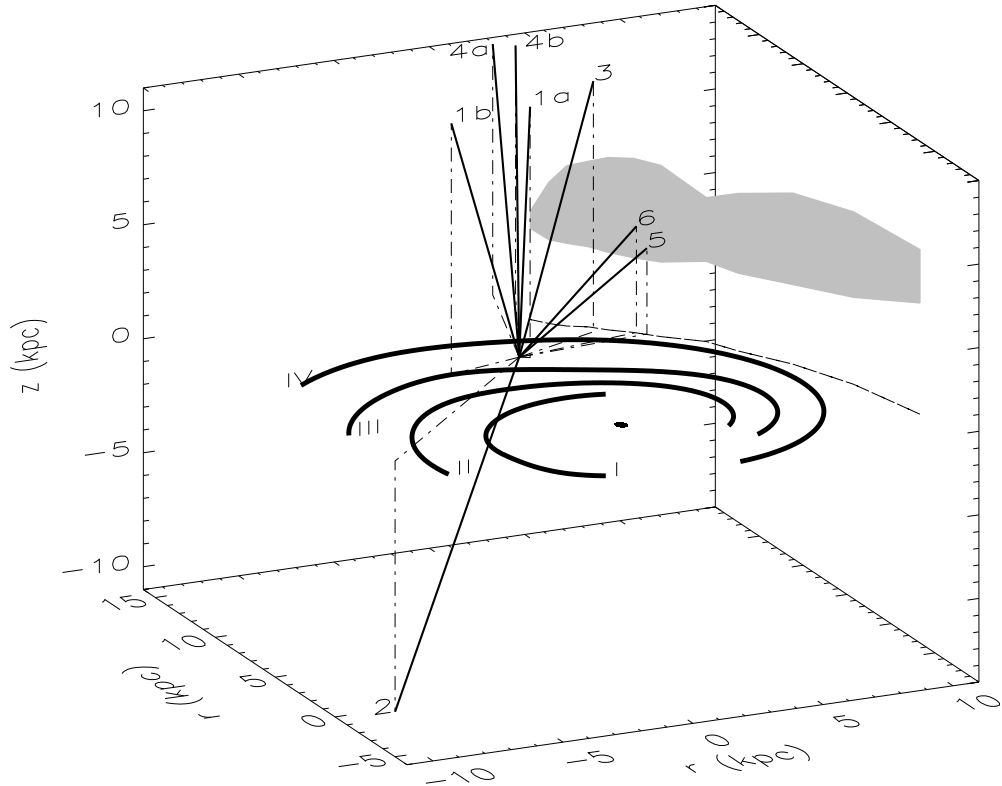}
\caption{\label {3d} 3D model of the Milky Way. The four major spiral arms are
shown as thick lines in the $z=0$ plane: I = Norma, II = Scutum-Crux, III =
Sagittarius-Carina, IV = Perseus. The light grey area represents a model for HVC
Complex C. The projection of Complex C into the Galactic disk is shown as a
dashed line. The \ion{O}{6} emission sight lines are shown as solid, straight
lines with their projections into the disk shown as dash-dotted lines: 1a, b =
\citet{dixon}; 2 = \citet{sh01}; 3 = \citet{sh02a}; 4a, b = \citet{wel}; 5 =
SL5; 6 = SL6. All sight lines are cut off at $|z|=11$\,kpc, except sight lines 5
and 6, the only sight lines intersecting HVC Complex C and thus cut off at the
Complex C surface.}
\end{figure}

Figure \ref{3d} shows a schematic picture of the Milky Way. The positions of the
four major spiral arms are adopted from \citet{tc}. The model of HVC Complex C
is based on \ion{H}{1} contours \citep{wak01} assuming a roughly constant
thickness and $z$ height for Complex C. No assumptions for the depth of Complex
C are made. All \ion{O}{6} emission sight lines observed with {\em FUSE} are
shown. While the other sight lines have Galactic latitudes of about 60$\degr$ or
more (except for sight line 2, hereafter SL2), SL5 has a latitude of only about
40$\degr$, thus probing a different part of the Milky Way.

SL5 intersects HVC Complex C at its southern boundary. Due to the large
extinction of Complex C, it is unlikely that the observed \ion{O}{6} emission
originates in gas beyond these HVCs, i.e. the location of Complex C can be
considered an upper limit on the distance to the \ion{O}{6} emitting gas. A
(weak) lower limit on the distance to Complex C at $l\approx 90\degr$ is 6\,kpc
\citep{wak01}. If the source of the \ion{O}{6} emission were the surface of
infalling clouds, we could expect \ion{O}{6} emission at the same high
velocities as measured in the Leiden-Dwingeloo \ion{H}{1} Survey in this
direction. The \ion{H}{1} HVC component is observed at
$v_{\rm LSR}=-115$\,km\,s$^{-1}$ (B. Wakker 2002, private communication). The
SL5 data show no \ion{O}{6} counterpart at this velocity. The 1$\sigma$ upper
limit for HVC \ion{O}{6} is 300\,LU. We conclude that the observed \ion{O}{6}
emission does not come from infalling HVCs.

\begin{figure}
\plotone{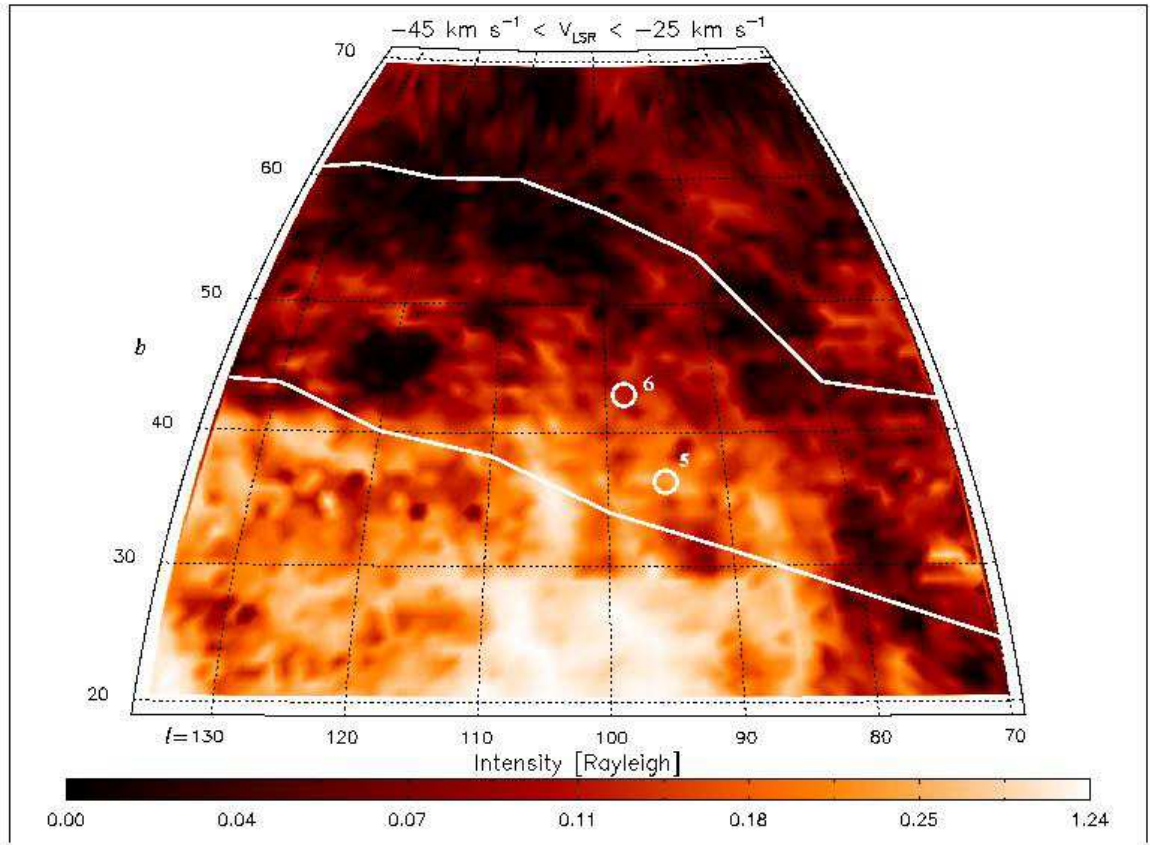}
\caption{\label{wham} H$\alpha$ image with {\em FUSE} positions and HVC Complex
C boundaries marked. The H$\alpha$ emission is integrated over the velocity
range from $-45$ to $-25$\,km\,s$^{-1}$; the intensity scale is not linear
(courtesy of M. Haffner \& G. Madsen). Overlaid in white are the faintest
contours of HVC Complex C \citep{wak01}. The positions of the SL5 and SL6 {\em
FUSE} sight lines are marked as circles. Note that the circles are not scaled to
the size of the LWRS aperture. The brightest H$\alpha$ emission at the bottom of
this map ($l=85\degr-110\degr$) is part of a group of strong filaments extending
from the disk into the halo.}
\end{figure}

The SL5 sight line passes $\sim$3\,kpc above the Perseus arm (Fig. \ref{3d}).
The velocity of the Perseus arm in the direction of SL5 ($l=95\fdg4$) is
$-70\pm10$\,km\,s$^{-1}$ \citep{kep}. Assuming corotation of associated gas at
higher altitudes, the Perseus arm velocity projected onto sight line SL5 at
$b=36\fdg1$ is $-58\pm8$\,km\,s$^{-1}$. This is consistent with the velocities
observed in the SL5 spectrum and implies that the \ion{O}{6} emitting gas has no
velocity component perpendicular to the disk. Figure \ref{wham} shows a map of
H$\alpha$ emission from the Wisconsin H-Alpha Mapper (WHAM) Survey (courtesy of
M. Haffner \& G. Madsen). A group of strong H$\alpha$ filaments extends between
$l=85\degr$ and $110\degr$ from the disk into the halo indicating an outflow of
gas. Sight line SL5 intersects an H$\alpha$ filament above this region of
strongest H$\alpha$ emission. The WHAM spectra closest to our SL5 pointing show
H$\alpha$ emission at velocities ranging from $-50$ to $+20$\,km\,s$^{-1}$ (M.
Haffner 2002, private communication). The H$\alpha$ filaments are most
pronounced in the velocity range from $-45$ to $-25$\,km\,s$^{-1}$ at higher
latitudes and $-25$ to 0\,km\,s$^{-1}$ closer to the disk. This suggests that
the outflow velocity decreases with $z$ in H$\alpha$. A simple Galactic rotation
model (a corotating halo with $v=220$\,km\,s$^{-1}$) places gas with the
observed \ion{O}{6} velocity at $z\approx 5.5$\,kpc and at a distance of $\sim
9.3$\,kpc from the Sun along sight line SL5 in the neighborhood of the Perseus
arm H$\alpha$ filaments. The similar velocities of these filaments, the Perseus
arm, and the \ion{O}{6} emitting gas suggest that these three components are
associated with each other: The shock-heated gas ejected by supernova explosions
in the Perseus arm and observed in SXR emission \citep{snow} is cooling down or
mixing with cooler gas and emitting the \ion{O}{6} photons that we see. The
\ion{O}{6} emitting gas is on the verge of falling back to the disk, while the
underlying outflow supplies new, hot gas.

The measured intensities in the SL5 data ($1600\pm300$\,LU at 1032\,\AA\ and
$800\pm300$\,LU at 1038\,\AA) are lower than all previously measured \ion{O}{6}
emission line intensities (2000--3300 and 1100--2000\,LU, respectively). All the
previous sight lines point toward high Galactic latitudes inside what
\citet{wel} call the Local Interstellar Chimney. Only SL2 and SL5 (both at
$|b|\approx 40\degr$) appear to intersect the walls of the Local Bubble and the
Chimney (see Sfeir et al. 1999 for contours). A new study by \citet{sh02b}
yields 2$\sigma$ upper limits of 420\,LU (1032\,\AA) and 540\,LU (1038\,\AA) for
\ion{O}{6} emission originating in the Local Bubble, intensities much lower than
any of the \ion{O}{6} detections. This suggests that a significant fraction of
the observed \ion{O}{6} emission along SL2 and SL5 originates beyond the Local
Bubble. The lower intensities in the SL5 data can be explained by more
extinction due to the larger distance to the \ion{O}{6} emitting region, if it
in fact lies above the Perseus arm as suggested above. Although the \ion{O}{6}
intensities differ by a factor of about 2 between SL2 and SL5, the SXR fluxes
are similar ($930\pm60$ and $944\pm15 \times
10^{-6}$\,counts\,s$^{-1}$\,arcmin$^{-2}$, respectively; Snowden et
al.~1997\footnote{Values obtained using the NASA HEASARC X-Ray Background Tool
at http://heasarc.gsfc.nasa.gov/cgi-bin/Tools/xraybg/xraybg.pl}). It is not
clear whether most of the observed SXR emission comes from the same region as
the \ion{O}{6} emission.

The correlations among the velocities of spiral arms, H$\alpha$ filaments, and
\ion{O}{6} emitting gas that we find for SL5 are not present for any of the
previously-observed \ion{O}{6} emission sight lines. For example, the spiral arm
and \ion{O}{6} velocities along SL2 are consistent only if the \ion{O}{6}
emitting gas has an outflow velocity of $100-170$\,km\,s$^{-1}$ at $z=1-9$\,kpc.
No H$\alpha$ filaments or other signs of outflows were found near the
previously-observed sight lines. (Note: SL2 points toward a region not observed
by the WHAM survey.) The corotating halo model would place the \ion{O}{6}
emitting gas $10-20$\,kpc above the plane (sight lines 2, 4a, 4b) or even higher
($>600$\,kpc and $>70$\,kpc for sight lines 1a and 1b); sight line 3 yields
positive \ion{O}{6} velocities, whereas the model requires negative velocities.
We therefore conclude that SL5 is the only sight line for which a reasonable and
consistent explanation of the \ion{O}{6} emission and its parameters can be
found, and that SL5 and possibly SL6 probe halo gas that originated in the
underlying spiral arm.

\section{CONCLUSIONS}

We have observed Galactic \ion{O}{6} emission along two low-latitude sight
lines, only one of which yields reliable measurements. Based on the observed
velocities of the \ion{O}{6} emitting gas, the H$\alpha$ filaments, and the
underlying spiral arm, we conclude that the \ion{O}{6} emission seen along sight
line SL5 (and probably SL6) originates in the halo about $\sim3$\,kpc above the
Perseus arm. Our observations are consistent with a Galactic fountain scenario
in which gas is heated and ejected by supernova explosions in the spiral arms,
becomes boyant in the colder disk gas, and rises into the halo where it cools
down and sooner or later falls back to the disk. We find no \ion{O}{6}
counterpart to the HVC component observed in \ion{H}{1}.

\acknowledgements

This research is supported by NASA contract NAS5-32985 to the Johns Hopkins
University. The authors thank L. M. Haffner and G. J. Madsen for creating the
WHAM map. The Wisconsin H-Alpha Mapper is funded by the National Science
Foundation. The authors are also thankful to B. P. Wakker for his comments on
HVC Complex C and the \ion{H}{1} line profiles. This work made use of the NASA
Extragalactic Database (NED) and the NASA Astrophysics Data System (ADS).

\clearpage

\begin{deluxetable}{llccccc}
\tablewidth{0pt}
\tablecaption{\label{obs} OBSERVATIONS}
\tablehead{\colhead{Position} & \colhead{Program} & \colhead{$\alpha$} &
\colhead{$\delta$} & \colhead{$l$} & \colhead{$b$} & \colhead{Exposure Time} \\
\colhead{} & \colhead{ID} & \colhead{(h m s)\tablenotemark{a}} &
\colhead{(d m s)\tablenotemark{a}} & \colhead{(deg)\tablenotemark{a}} &
\colhead{(deg)\tablenotemark{a}} & \colhead{(s)\tablenotemark{b}}}
\startdata
SL5 & S4054801 & 16 59 38.39 & 65 00 43.83 & 95.4 & 36.1 & 217352/94829 \\
SL6 & S4056101 & 15 46 02.23 & 65 01 37.87 & 99.3 & 43.3 & 31553/13114 \\
\enddata
\tablenotetext{a}{Coordinates are J2000}
\tablenotetext{b}{Day-plus-night/night only}
\end{deluxetable}

\begin{deluxetable}{lccccc}
\tablewidth{0pt}
\tablecaption{\label{meas} EMISSION LINE PARAMETERS FOR SL5}
\tablehead{\colhead{Line} & \colhead{$I_{\rm tot}$\tablenotemark{a}
\tablenotemark{b}} &
\colhead{$I_{\rm tot}$\tablenotemark{b} \tablenotemark{c}} &
\colhead{FWHM}\tablenotemark{b} & \colhead{$\lambda_{\rm cen}$} &
\colhead{$v_{\rm LSR}$} \\
\colhead{} & \colhead{(ph\,s$^{-1}$\,cm$^{-2}$\,sr$^{-1}$)} &
\colhead{(ph\,s$^{-1}$\,cm$^{-2}$\,sr$^{-1}$)} &  \colhead{(km\,s$^{-1}$)} &
\colhead{(\AA)} & \colhead{(km\,s$^{-1}$)}}
\startdata
\ion{O}{6}\,$\lambda$1031.926 & $1600\pm300$ & $1680\pm90$ & $75\pm3$ &
$1031.7\pm0.1$ & $-50\pm30$ \\
\ion{O}{6}\,$\lambda$1037.617 & $800\pm300$ & $900\pm700$ & $22\pm15$ &
$1037.40\pm0.02$ & $-44\pm6$ \\
\ion{C}{2}$^\ast\,\lambda$1037.02 & $1000\pm300$ & $1300\pm900$ & $40\pm20$ &
$1036.77\pm0.04$ & $-53\pm12$ \\
\enddata
\tablenotetext{a}{Integrated intensity using photon-counting method}
\tablenotetext{b}{Systematic uncertainties are not included in error estimates.}
\tablenotetext{c}{Integrated intensity using minimum $\chi^2$ method}
\end{deluxetable}

\clearpage

\end{document}